\documentclass[12pt]{article}  
\usepackage{hyperref}
\usepackage{color}
\usepackage{ulem}
\usepackage{epsfig,amsfonts,amssymb}

\usepackage[english]{babel}
\usepackage{amsmath,amssymb}
\usepackage{amsfonts}
\usepackage{mathrsfs}
\usepackage{pgf,pgfarrows}
\usepackage{epic}
\usepackage{eepic}
\usepackage{color}
\usepackage[all,cmtip,line]{xy}
\usepackage{graphicx}

\newcommand{\comma}{\;\; ,}
\newcommand{\period}{\;\; .}
\newcommand{\eq}{\; = \;}
\newcommand{\sep}{\;\; , \;\;}

\newcommand{\be}{\begin{equation}}
\newcommand{\bd}{\begin{displaymath}}
\newcommand{\ee}{\end{equation}}
\newcommand{\ed}{\end{displaymath}}
\newcommand{\ba}{\begin{eqnarray}}
\newcommand{\ea}{\end{eqnarray}}

\renewcommand{\i}{{\rm i}}
\newcommand{\e}{{\rm e}}

\newcommand{\bl}{\color{\ctnclr} }

\renewcommand{\i}{{\rm i}}

\renewcommand{\theequation}{\arabic{equation}}
\renewcommand{\theequation}{\arabic{equation}}
\newcommand{\ctnclr}{blue}

\title{Onsager and Kaufman's calculation of the spontaneous 
magnetization of the Ising model: II }

\author{ R.J. Baxter\\
{\protect \small  Mathematical
Sciences Institute}\\
{\protect \small  The Australian National University,
 Canberra, A.C.T. 0200, Australia}\\
 }

\date{}

\begin{document}


\maketitle

\definecolor{magenta}{rgb}{0.5,0,0.5}

\definecolor{red}{rgb}{0.8,0,0}

\definecolor{green}{rgb}{0,0.5,0}

\definecolor{blue}{rgb}{0,0,0.8}

\vspace{2cm}

\definecolor{black}{rgb}{0.4,0.4,0.4}

\definecolor{brown}{rgb}{0.4,0.2,0.2}

 \setlength{\unitlength}{1pt}

 \abstract{In 2011 I reviewed the calculation by Onsager and Kaufman of 
 the spontaneous magnetization of the square-lattice Ising model, which 
 Onsager announced in 1949 but never published. I have recently been alerted to further
 original papers that bear on the subject. It is quite clear that the draft 
 paper on which I relied was indeed written by Onsager, who was working 
 on the problem with Kaufman, and that they had two 
 derivations of the result.

 \vspace{5mm}


\setlength{\parskip}{0mm}


 \section*{Introduction}
 


\setcounter{equation}{0}
\renewcommand{\theequation}{\arabic{equation}}

In my original paper of 2011 on Onsager and Kaufman's work on the spontaneous 
magnetization $M_0$ of the Ising model\cite{baxter2011}, I presented a draft paper that 
I had been given some years earlier by John Stephenson , who had received it in about 
1965 from Ren Potts. It bears the hand-written names of Onsager and Kaufman. I 
presented it both as a directly scanned 
version and (for clarity) a transcript: I shall refer to it herein as OK.

I was concerned with the puzzle of why Onsager had announced in 1949 that he and 
Kaufman had a proof of the result for spontaneous magnetization of the Ising model, but 
had never published that proof.
I was aware of various writings that bore on the problem, including those in 
the Onsager archive in Trondheim, at

\setlength{\parskip}{0.6mm}


\noindent {\footnotesize {\color{blue}
{\url{http://www.ntnu.no/ub/spesialsamlingene/tekark/tek5/arkiv5.php}}}}

\noindent In particular, I was aware of the section ``Selected research material and 
writings" and the sub-sections 9.94 -- 10.104 headed ``Ising Model''. I quoted 
sub-section 9.97, which contains material relevant to the calculation of the 
spontaneous magnetization.

\setlength{\parskip}{0mm}

Professor Percy Deift of the Courant Institute in New York  has recently alerted me to another 
sub-section of the Onsager archive, namely  17.120 -- 17.129, headed simply ``Writings'' and 
containing three documents. I was remiss not to have found this earlier as it is very relevant. 
Here I briefly review my previous comments in the light of these documents.

 \section*{Onsager and Kaufman's two methods}

In \cite{baxter2011} I quoted Onsager as saying that he had reduced the problem 
to one of calculating a 
particular $k$ by $k$  Toeplitz determinant $D_m$ in the limit  $m \rightarrow \infty$, 
and that he solved the problem in two ways, the first by using generating functions 
and using a parametrization in terms of elliptic functions to reduce it to an integral 
equation problem with a kernel that was the sum of two parts, one a function of the 
difference of the two parameters, the other 
a sum.\cite[p. 11]{Onsager1971a}, \cite[eqn 2.7] {baxter2011})

In the second way, Onsager found a formula for the $m \rightarrow \infty$ limit of 
$\Delta_r$ for a general class of Toeplitz determinants $\Delta_r$.  It is contained in 
the draft paper OK and 
in a letter from Onsager to Kaufman. We outline the method below.

\subsection*{Summary of the second method}

The draft paper OK  is concerned with the evaluation of an $r$ by $r$  Toeplitz 
determinant $\Delta_r$, with entries  $c_{j-i}$ in row $i$ and column $j$, 
for $1 \leq i,j \leq r$. Let $f(z)$ be the generating function with 
coefficients  $c_j$:

\be  f(z)  \eq \sum_{n=-\infty}^{\infty}  c_n \, z^n  \period \ee
The paper shows algebraically that if, for a finite number of $\alpha_j, \beta_k$,
\be \label{formf}
 f(z) \eq \frac{\prod_j{(1-\alpha_j z)^{m_j}}}{\prod_k{(1-\beta_k  z^{-1} )^{n_k}} }
 \comma \ee
then
\be \label{del} \Delta_r \eq \prod_j \prod_k (1-\alpha_j \beta_k )^{m_j n_k}  \ee
provided the $m_j$,  $n_k$ are non-negative integers, $|\beta_k| <1 $ and 
$r  \geq \sum_k n_k $.

Further,  on pages 21- 24 of sub-section 9.97 of the Onsager archive, there is a letter from 
Onsager to Kaufman. He 
defines   $\Delta_r$ (or  $\Delta_k$) as above and sets
\be \e^{\eta_{+} } \eq \prod_j{(1-\alpha_j z)^{m_j}} \sep \e^{-\eta_{-}}  \eq 
\prod_k{(1-\beta_k  z^{-1} )^{n_k}} \period \ee
He then quotes the result (\ref{del}) (in this letter Onsager negates the $n_k$) and states 
that this implies
\be \label{etaform}
\log \Delta_{\infty} \eq \frac{\i}{2 \pi} \,  \int_{\omega = 0}^{2 \pi}  \, \eta_{+} \, d \eta_{-} 
( \omega )  \ee
where Onsager takes $z$ above to be $\e^{\i \omega}$.

If we define $b_n$ so that 
\be \log f(z) = \sum_{n= -\infty}^{\infty} b_n z^n \comma \ee
then $b_0 = 0$, 
\be \eta_{+} \eq \sum_{n=1}^{\infty} b_n z^n \sep  \eta_{-} \eq
 \sum_{n=1}^{\infty} b_{-n}  z^{-n} \ee 
and we can write (\ref{etaform}) as
\be \label{res}
 \log \Delta_{\infty} \eq  \sum_{n=1}^{\infty} n b_n b_{-n}  \period \ee

This reasoning depends on $f(z)$ having the form  (\ref{formf}), with integer $m_j, n_k$.
However, Onsager obviously realises that the result (\ref{etaform}) - (\ref{res})  must have 
greater validity.\footnote{ If $\log f(z)$ is analytic on the unit circle $|z| = 1$,
then the sum in (\ref{res}) converges and can be approximated to any accuracy by a finite 
truncation of the values of $n$. Since  (\ref{formf})
is sufficiently general to fit any finite number of the $b_n$, this suggests that (\ref{res}) should 
apply to any function $f(z)$ analytic on the unit circle. One needs to worry about convergence. }
He begins his letter to Kaufman by saying  ``This is to let you know that I have found a general 
formula for the value of an infinite recurrent determinant". He applies this formula to the 
determinant needed for the spontaneous 
magnetization and obtains the  now well-known result $M_0  = (1- k^2)^{1/8}$.

He and Kaufman preferred this method, but were working on ``how to to fill out the holes in the 
mathematics and show the epsilons and deltas and all of that''\cite[p. xxiii]{Onsager1971b},   
when the mathematicians Kakutani and 
Szeg{\H o} became aware of their work and  ``got there first''.\cite[p. 12]{Onsager1971a}   In 
fact Szeg{\H o}'s paper, in which he gives the result (\ref{res}) for the case 
when $b_{-n} = b^*_n$,  
did not appear until 1952\cite{Szego1952} and did not refer to the spontaneous magnetization 
problem. The first publication of a  proof of the formula for $M_0$ was  by C.N. Yang, also  
in 1952.\cite{Yang1952}  The general formula (\ref{res}) appears in  a 1963 paper by 
Montroll, Potts and Ward.\cite[eqn. 68] {MPW}

\section*{Further material}

Sub-section 17.121 of the Onsager archive (the one of which I was unaware in 2011) is 
entitled  ``Crystal Statistics. IV. Long-range order in a binary 
crystal'' (with B. Kaufman). It contains three documents:

{\bf 1.}  Pages 1  -- 4 is the letter from Onsager to Kaufman mentioned above, dated April 12, 
1950, giving a fomula for the determinant of a  general $k$ by $k$  Toeplitz determinant in 
the limit  $k \rightarrow \infty$.
This is also on pages 21 - 24 of sub-section 9.97 and is the letter quoted above. I refer to it  
section 5 (sub-section 1) of my 2011 
paper.

{\bf 2.} Pages 5 --12 contain the draft paper OK I presented in \cite{baxter2011}. I was working 
from a photo-copy of a photo-copy of a carbon copy.
The version on the archive is clearer and appears to be from the original type-script (or at 
least a better carbon). My copy is interesting in that it contains hand-written corrections 
and additions, probably by Onsager or Kaufman themselves. The fact that the paper is 
on the Onsager archive is a clear indication that it is indeed by Onsager, who was working 
in collaboration with Kaufman.

{\bf 3.} Pages 13 - 22 contain the draft of a paper headed  ``Crystal Statistics. IV. Long-range 
order in a binary crystal'' by Lars Onsager and Bruria Kaufman. This appears to be giving 
Onsager's first method. It is unfinished, but is certainly  leading towards an integral equation 
with a kernel with difference and sum properties. In Appendix A of \cite{baxter2011} I give a 
calculation of $M_0$ that involves such a kernel. Onsager and Kaufman's draft begins 
differently, but appears to be heading in a similar direction.

So Onsager and  Kaufman did  indeed have two ways of proving the formula for $M_0$.  They 
turned their attention to other problems when the mathematicians became interested in their 
preferred method, so never published either way of solving the problem.

\end{document}